# Neural network-based image reconstruction in swept-source optical coherence tomography using undersampled spectral data

Yijie Zhang[1,2,3,†], Tairan Liu[1,2,3,†], Manmohan Singh[4], Yilin Luo[1,2,3], Yair Rivenson[1,2,3], Kirill V. Larin[4,5], and Aydogan Ozcan[1,2,3,6,*]

[1]Electrical and Computer Engineering Department, University of California, Los Angeles, CA 90095, USA
[2]Department of Bioengineering, University of California, Los Angeles, CA 90095, USA
[3]California NanoSystems Institute, University of California, Los Angeles, CA 90095, USA
[4]Department of Biomedical Engineering, University of Houston, Houston, TX 77204, USA
[5]Department of Molecular Physiology and Biophysics, Baylor College of Medicine, University of Houston, Houston, TX 77204, USA
[6]Department of Surgery, David Geffen School of Medicine, University of California, Los Angeles, CA 90095, USA
[†]Equally contributing authors
*Corresponding author: Aydogan Ozcan ozcan@ucla.edu

**Abstract:**
Optical Coherence Tomography (OCT) is a widely used non-invasive biomedical imaging modality that can rapidly provide volumetric images of samples. Here, we present a deep learning-based image reconstruction framework that can generate swept-source OCT (SS-OCT) images using undersampled spectral data, without any spatial aliasing artifacts. This neural network-based image reconstruction does not require any hardware changes to the optical set-up and can be easily integrated with existing swept-source or spectral domain OCT systems to reduce the amount of raw spectral data to be acquired. To show the efficacy of this framework, we trained and blindly tested a deep neural network using mouse embryo samples imaged by an SS-OCT system. Using 2-fold undersampled spectral data (i.e., 640 spectral points per A-line), the trained neural network can blindly reconstruct 512 A-lines in ~6.73 ms using a desktop computer, removing spatial aliasing artifacts due to spectral undersampling, also presenting a very good match to the images of the same samples, reconstructed using the full spectral OCT data (i.e., 1280 spectral points per A-line). We also successfully demonstrate that this framework can be further extended to process 3× undersampled spectral data per A-line, with some performance degradation in the reconstructed image quality compared to 2× spectral undersampling. This deep learning-enabled image reconstruction approach can be broadly used in various forms of spectral domain OCT systems, helping to increase their imaging speed without sacrificing image resolution and signal-to-noise ratio.

## Introduction

Optical coherence tomography (OCT) is a non-invasive imaging modality that can provide three-dimensional (3D) information of optical scattering properties of biological samples. The first generation of OCT systems were based on time-domain (TD) imaging[1], using mechanical path-length scanning. However, the relatively slow data acquisition speed of the early TDOCT systems partially limited their applicability for *in vivo* imaging applications. The introduction of the Fourier Domain (FD) OCT techniques[2,3] with higher sensitivity[4,5] has contributed to a dramatic increase in imaging speed and quality[6]. Modern FDOCT systems can routinely achieve line rates of 50-400 kHz[7–12] and there have been recent research efforts to further improve the speed of A-scans to tens of MHz[13,14]. Some of these advances employed hardware modifications to the optical set-up to improve OCT imaging speed and quality, and focused on e.g., improving the OCT system design, including improvements in high speed sources[13,15,16], also opening up new applications such as single-shot elastography[17] and others[18–20].

Recently, we have experienced the emergence of deep-learning-based image reconstruction and enhancement methods[21–23] to advance optical microscopy techniques, performing e.g., image super resolution[23–28], autofocusing[29–31], depth of field enhancement[32–34], holographic image reconstruction and phase recovery[35–38], among many others[39–42]. Inspired by these applications of deep learning and neural networks in optical microscopy, here we demonstrate the use of deep learning to reconstruct swept-source OCT (SS-OCT) images using undersampled spectral data points. Without the need to perform any hardware modifications to an existing SS-OCT system, we show that a trained neural network can rapidly process undersampled spectral data and match, at its output, the image quality of standard SS-OCT reconstructions of the same samples that used 2-fold more spectral data per A-line.

A major challenge in reducing the number of spectral data points in an OCT system is that the raw data acquired per A-line do not represent a smooth function with natural continuity between the neighboring pixels/frequencies, and instead exhibit rapid changes as a function of the wavelength. In our approach, we first reconstructed each A-line with 2-fold less spectral data (eliminating every other frequency), which resulted in severe spatial aliasing artifacts. We then trained a deep neural network to remove these aliasing artifacts that are introduced by spectral undersampling, matching the image reconstruction results that used all the available spectral data points. To demonstrate the success of this deep learning-based OCT image reconstruction approach, we used an SS-OCT[3] system to image murine embryo samples. The trained neural network successfully generalized, and removed the spatial aliasing artifacts in the reconstructed images of new embryo samples that were never seen by the network before. We further extended this framework to process 3× undersampled spectral data per A-line, and showed that it can be used to remove even more severe aliasing artifacts that are introduced by 3× spectral undersampling, although at the cost of some degradation in the reconstructed image quality compared to 2× spectral undersampling results.

In addition to overcoming spectral undersampling related image artifacts, the inference time of the deep neural network is also optimized, achieving an average image reconstruction time of 6.73 ms for 512 A-lines, processed all in parallel using a desktop computer; this inference time is further improved to ~1.69 ms by simplifying the neural network architecture.

We believe that this deep learning-based OCT image reconstruction method has the potential to be integrated with various swept-source or spectral domain OCT systems, and can potentially improve the 3D imaging speed without a sacrifice in resolution or signal-to-noise of the reconstructed images.

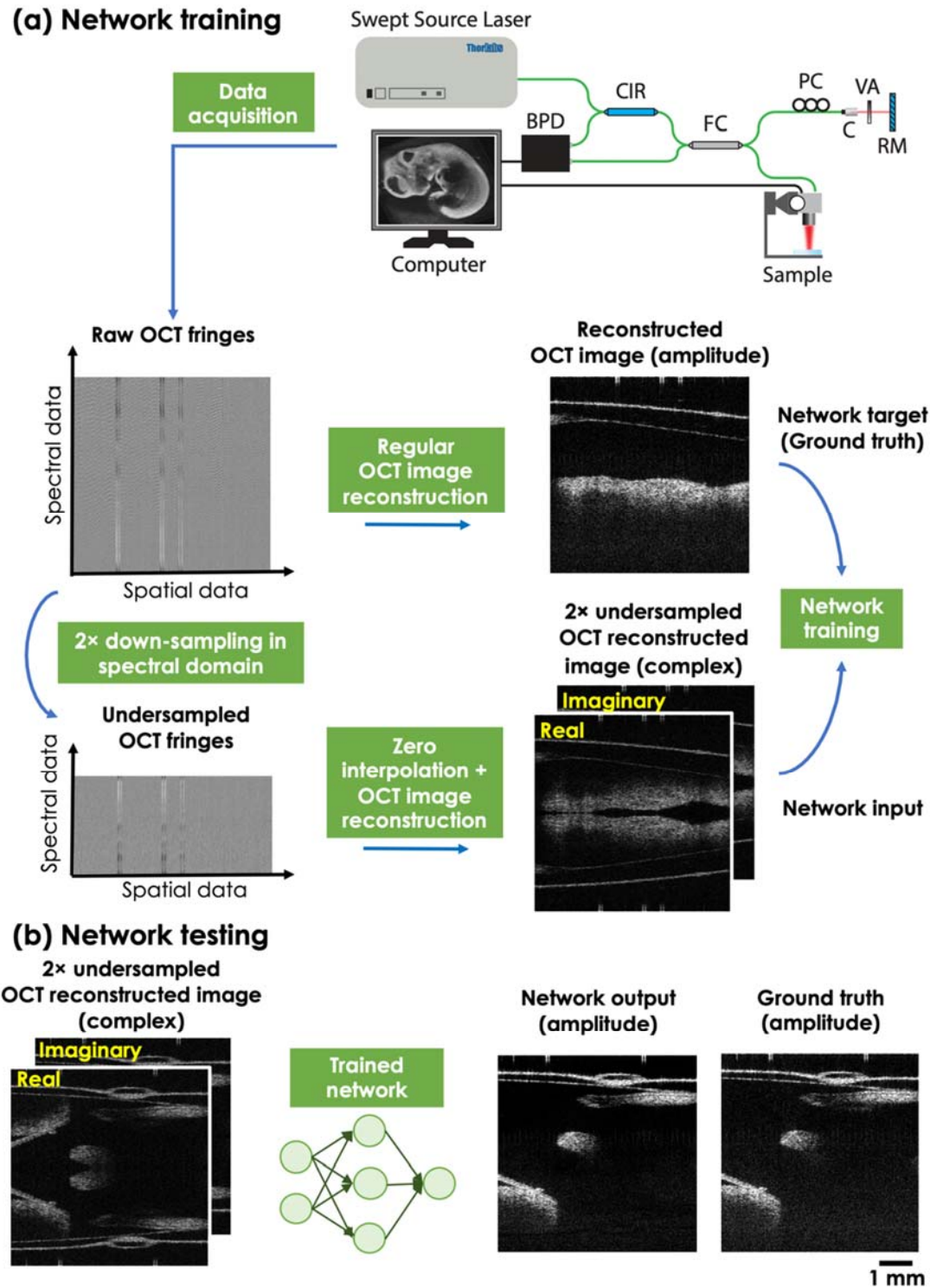

**Fig. 1 Schematic of the DL-OCT image reconstruction framework.** (a) Training phase of DL-OCT. Raw OCT fringes were captured by an SS-OCT system. The network target (ground truth) was generated by direct reconstruction of the original OCT fringes as detailed in the Materials and Methods. The network input was generated by 2-fold down-sampling of the spectral data for each A-line, zero interpolation, and reconstruction of the resulting fringes. (b) Testing phase of the DL-OCT. We pass the 2× undersampled OCT image (real and imaginary parts) through a trained network model to create an aliasing-free OCT image, matching the ground truth reconstruction that used all the spectral data points (see the Materials and Methods for details).

**Results**

To demonstrate the efficacy of this deep learning-based OCT image reconstruction framework, which we term DL-OCT, we trained and tested a deep neural network (see Materials and Methods section) using SS-OCT images acquired on mouse embryo samples. Our 3D image dataset consisted of eight different embryo samples, where five of them were used for training and the other three were used for blind testing. For each one of these embryo samples, 1000 B-scans (where each B-scan consists of 5000 A-lines, and each A-line has 1280 spectral data points) were collected by the SS-OCT system shown in Fig. 1(a); see Materials and Methods for more details. During the network training phase, the original OCT fringes per A-line were first reconstructed using a Fourier transform based image reconstruction algorithm to form the network's target (i.e., ground truth) images. Then, the same spectral fringes were 2× down-sampled (by eliminating every other spectral data point), zero interpolated, and reconstructed using the same Fourier transform based image reconstruction algorithm to form the input images of the network, each of which showed severe aliasing artifacts due to the spectral undersampling (see Figs. 1-2). Both the real and imaginary parts of these aliased OCT images were used as the network input, where only the amplitude channel of the ground truth was used for the target image during the training phase. After the network training process, which is a one-time effort, taking e.g., ~18 h using a desktop computer (see Materials and Methods section), the trained neural network successfully generalized and could reconstruct the images of unknown, new samples that were never seen by the network before, removing the aliasing related artifacts as shown in Fig. 1. Figure 2 further reports a detailed comparison of the network's input, output and ground truth images corresponding to different fields of view of mouse embryos, also quantifying the absolute values of the spatial errors made.

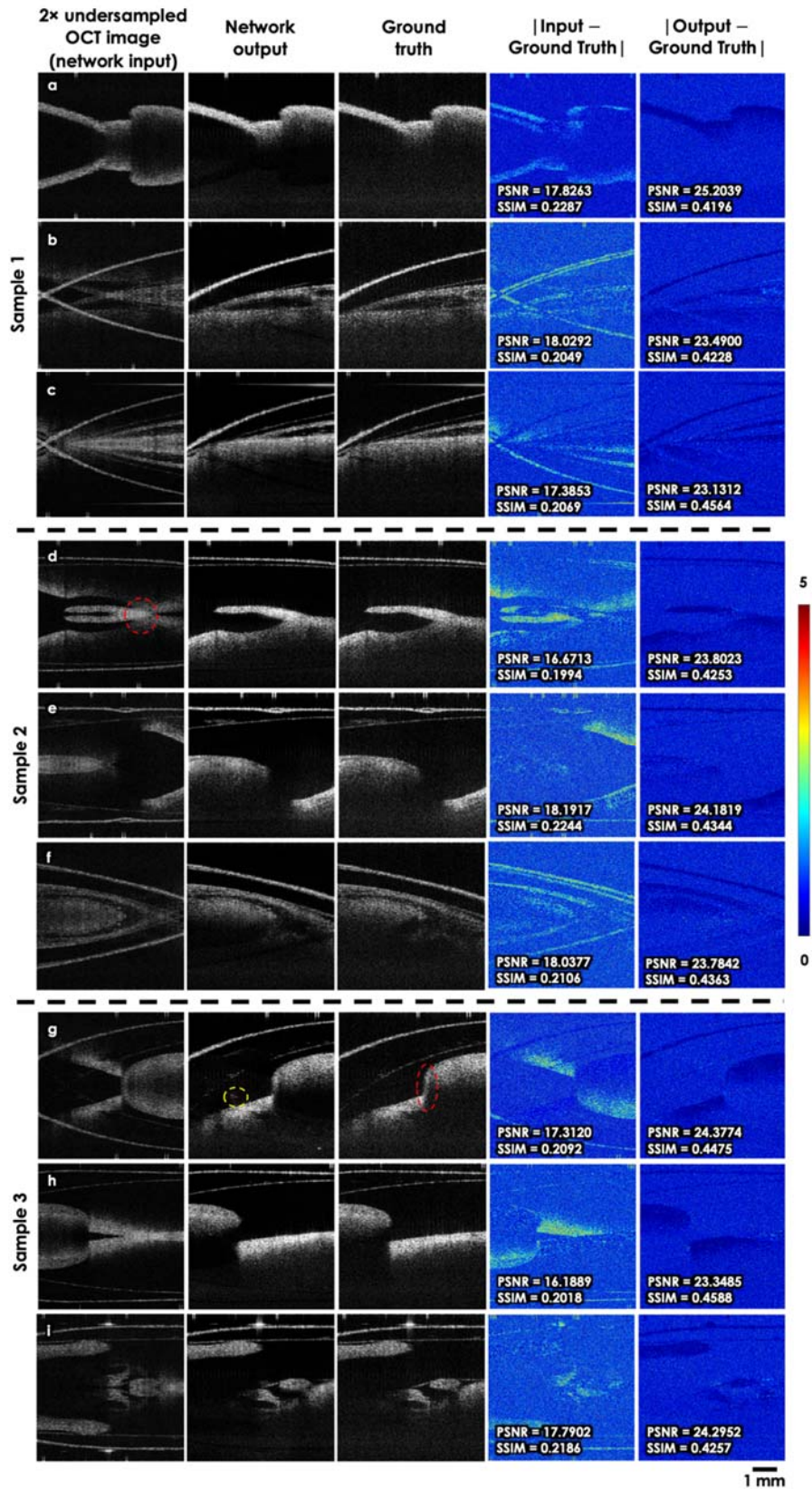

**Fig. 2 Blind testing performance of the DL-OCT framework.** The network input, output and ground-truth images of three mouse embryo samples (never seen by the network before) at different

fields-of-view are shown in the first three columns. The error maps of the input and output with respect to the corresponding ground-truth image are provided in 4th and 5th columns, respectively. Averaged peak signal to noise ratio (PSNR) and structural similarity index (SSIM) values were also computed and displayed for each one of these sample fields-of-view.

The reconstruction results reported in Figs. 1-2 clearly reveal that the trained network does not simply keep the connected upper part of the input image as the output. For example, in Fig. 2(g), the signal in the ground truth image crosses both the upper and the lower parts of the field-of-view, and in the red circled region, there is an abrupt change, breaking the horizontal connectivity of the image. The DL-OCT network learned to reconstruct the output images by utilizing a combination of the vertical morphological information exhibited in the target images and the special corrugated patterns caused by aliasing. In an OCT system, the illumination beam naturally forms an axially decaying pattern, where the surfaces or structural discontinuities usually have stronger signal than the internal structure of the sample[43]. This characteristic information was effectively captured by the neural network inference, as shown in for example Fig. 2(g). This also explains the occasional weak artifacts observed at the network output (see e.g., the yellow circled region in Fig. 2(g)) for features that lack detectable morphological information along the vertical axis. In general, the trained neural network uses both the vertical and horizontal information at the input image (within its receptive field) to remove various challenging forms of aliasing artifacts such as those emphasized with red color in Fig. 2d.

Next, to quantify the performance of DL-OCT image reconstructions, two quantitative metrics were calculated for 13,131 different test images: peak signal-to-noise ratio (PSNR) and the structural similarity index (SSIM) (see Materials and Methods for details). PSNR is a non-normalized metric which represents an estimation of the human perception of the image reconstruction quality. For images with pixels ranging from 0 to 1 with double precision (such as the test images in our framework), a 20 dB to 30 dB PSNR value is generally acceptable for noisy target images[44]. The SSIM, on the other hand, is a normalized metric that focuses more on image structure similarity between two images. This metric can take a value between 0 and 1 (where 1 represents an image that is identical to the target)[44]. Overall, compared to the target (ground truth) images that used all the spectral data points, the spectrally-undersampled *input* images with aliasing artifacts achieved a PSNR and an SSIM of 18.3320 dB and 0.2279, respectively, averaged over 13,131 test images. Both of these metrics were significantly improved at the network's output images, achieving 24.6580 dB and 0.4391, respectively, also averaged over 13,131 test images. Some examples of these image comparisons with the resulting PSNR and SSIM values are also reported in Fig. 2.

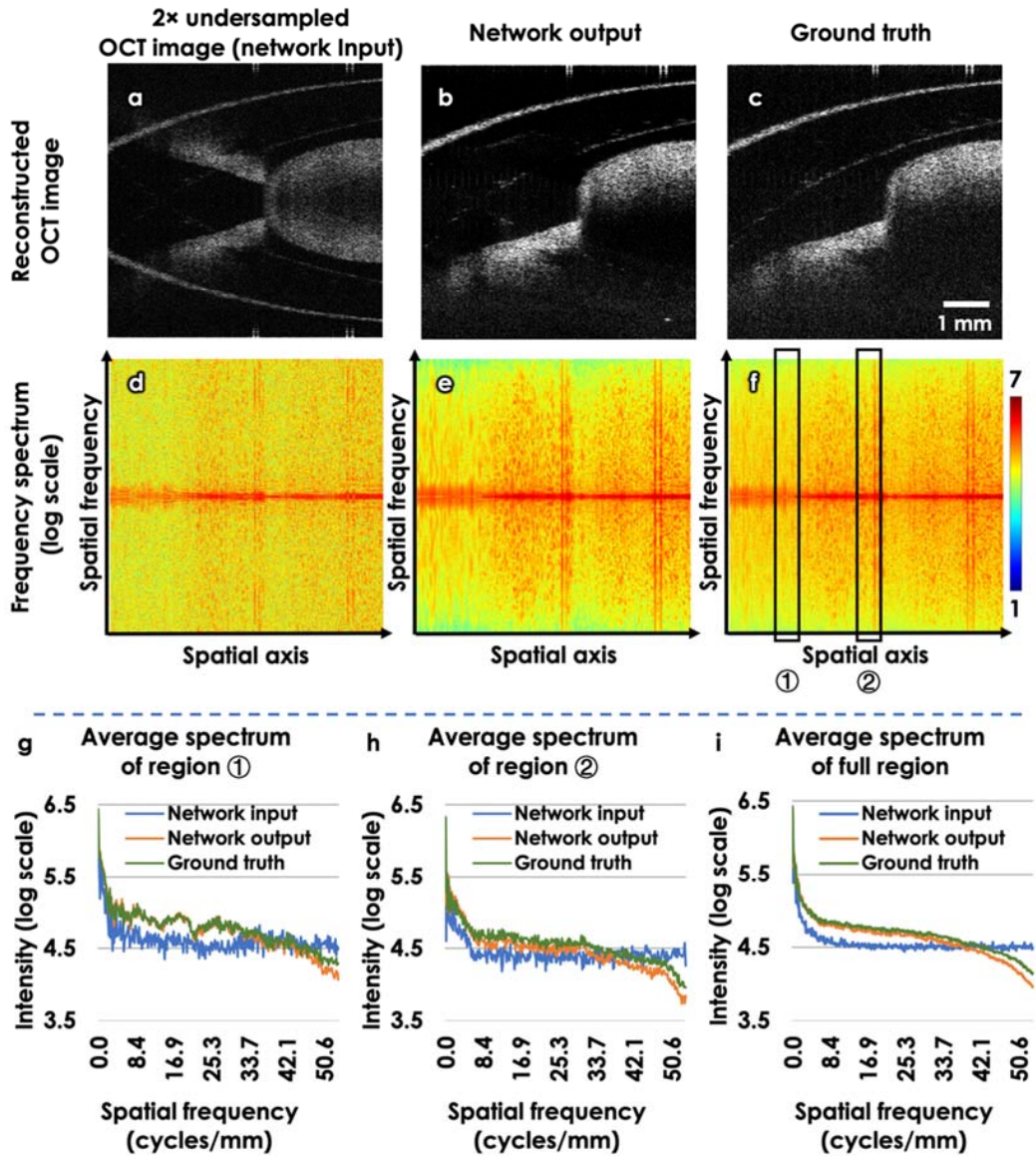

**Fig. 3 Frequency spectrum analysis of DL-OCT**. (a-c) SS-OCT images of a sample field-of-view, forming the network input, network output and ground truth, respectively. (d-f) Log-scaled spatial frequency spectra of (a-c) represented in spectral-spatial domain using 1D Fourier transform along the A-line direction of each image. (g-h) Averaged intensity of the spectral profile over two specific spatial regions (① and ② shown in (f)). (i) is the same as in (g-h), except that it is averaged over the entire spatial axis, shown in (d-f).

We also used spatial frequency analysis to further quantify our network inference results against the ground truth images. To perform this comparison, we converted the network input, output and ground truth images into spatial frequency domain by performing a 1D Fourier transform along the vertical axis (for each A-line). The results of this spatial frequency comparison for each A-line are shown in Figs. 3(d-f), which further reveal the success of the network's output inference, closely matching the spatial frequencies of the corresponding ground truth image. The quantitative comparison in Fig. 3(g-i) also demonstrates that the network output very well matches the ground truth images for both the low and high frequency parts of a sample.

**Discussion**

In our results reported so far, we used zero interpolation to pre-process the 2× undersampled spectral data per A-line, before generating the network's input image with severe spatial aliasing. Alternatively, zero-padding is another method that can be used to pre-process the undersampled spectral data for each axial line. However, other spectral interpolation methods such as the nearest neighbor, linear, or cubic interpolation may result in various additional artifacts due to the non-smooth structure of each spectral line. We performed a comparison of these different interpolation methods used to pre-process the same undersampled spectral data, the results of which are summarized in Fig. 4; in these results, each DL-OCT network was separately trained using the same undersampled spectral data, pre-processed using a different interpolation method. Among these interpolation methods, cubic interpolation was found to generate the most severe spatial artifacts at the network output. Both zero padding and zero interpolation methods shown in Fig. 4 consistently resulted in successful image reconstructions at the network output, removing aliasing artifacts observed at the input images, providing a decent match to the ground truth. On the contrary, other interpolation methods, such as cubic interpolation, introduced additional artifacts at the network output image (see e.g., the red circled region in Fig. 4c) due to the inconsistent interpolation of missing spectral data points at the input. To further quantify this comparison, we also calculated the SSIM and PSNR values between the network output images and the corresponding ground truth SS-OCT images for five different pre-processing methods (see Table 1). This quantitative analysis reported in Table 1 reveals that the zero interpolation method (presented in the Results section) achieves the highest PSNR and SSIM values for reconstructing SS-OCT images using 2-fold undersampled spectrum per A-line. It is also worth noting that the zero interpolation and zero padding methods achieve very close quantitative results, and significantly outperform the other spectral interpolation methods, including cubic, linear and nearest neighbor interpolation, as summarized in Table 1.

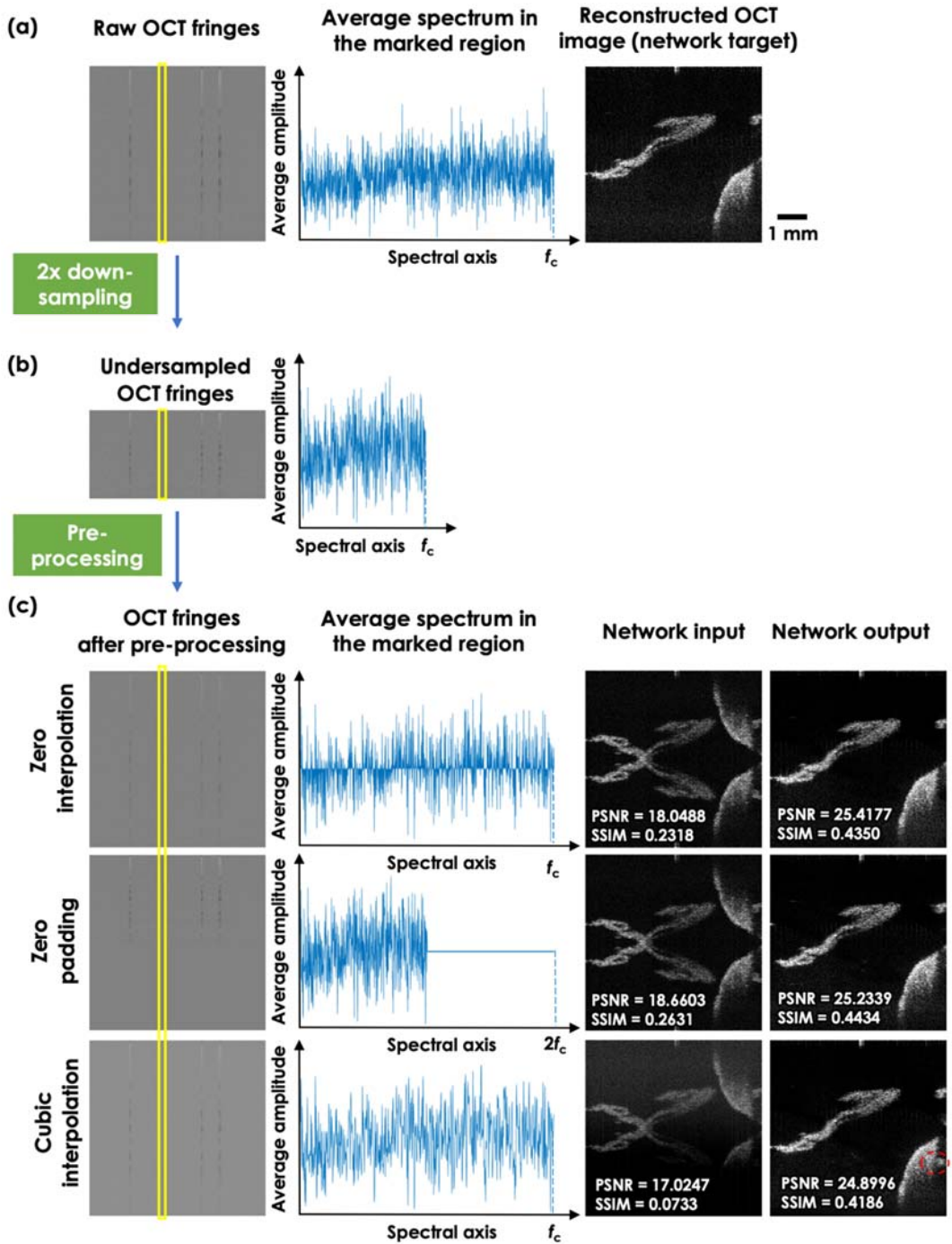

**Fig. 4 Comparison of different pre-processing methods for DL-OCT.** (a) Raw SS-OCT spectral fringes and the corresponding reconstructed OCT image (ground truth), where $f_c$ indicates the cut-off frequency of the spectral data. (b) 2-fold undersampled OCT fringes. (c) Undersampled OCT fringes that are pre-processed using different interpolation methods. Three separate neural networks were trained for each one of the pre-processing methods to generate the network outputs. PSNR and SSIM values are also displayed for each one of these fields-of-view.

**Table. 1 Comparison of PSNR and SSIM values between the network output images and the corresponding ground truth SS-OCT images for five different pre-processing methods. Also see Fig. 4.**

| Processing method | PSNR | | SSIM | |
| --- | --- | --- | --- | --- |
| | Average | Standard deviation | Average | Standard deviation |
| Zero interpolation (Results Section) | **24.6580** | 1.3225 | **0.4391** | 0.0320 |
| Zero-padding | 24.3447 | 1.8167 | 0.4378 | 0.0338 |
| Cubic interpolation | 24.3610 | 1.2398 | 0.4018 | 0.0291 |
| Linear interpolation | 24.4303 | 1.3493 | 0.4169 | 0.0343 |
| Nearest neighbor interpolation | 24.4763 | 1.3144 | 0.4135 | 0.0288 |

We also analyzed the inference speed of the trained DL-OCT network to reconstruct SS-OCT images with undersampled spectral measurements. For a batch size of 128 B-Scans, where each B-scan consists of 512 A-lines (with 640 spectral data points per A-line), the neural network is able to output a new OCT image in ~6.73 ms per B-scan using a desktop computer (see Fig. 5). This inference time can be further reduced with some simplifications made in the neural network architecture; for example, a reduction of the number of channels from 48 to 16 at the first layer of the neural network (Fig. 6) helped us reduce the average inference time down to ~1.69 ms per B-scan (512 A-lines), as shown in Fig. 5. With additional parallelization through the use of a larger number of GPUs, the inference speed per B-scan can be further improved to serve various applications that demand rapid reconstruction of 3D samples.

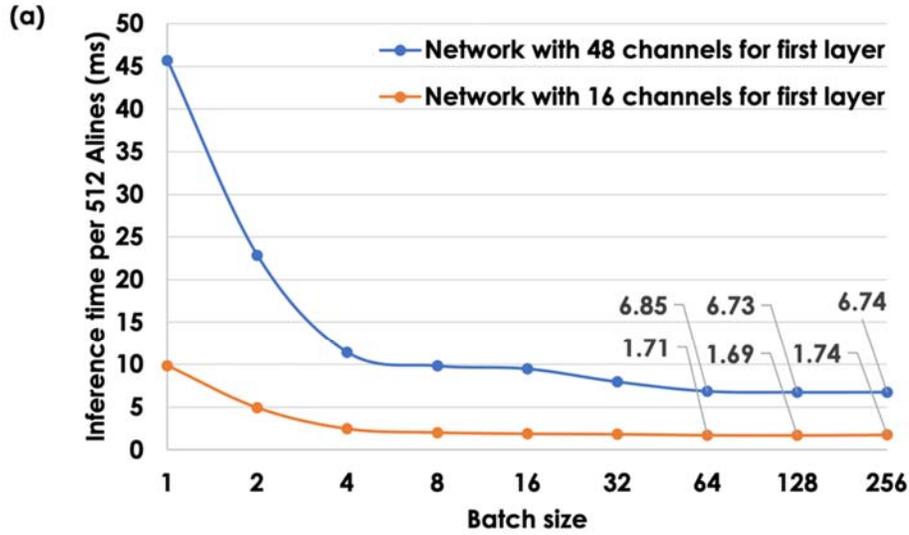

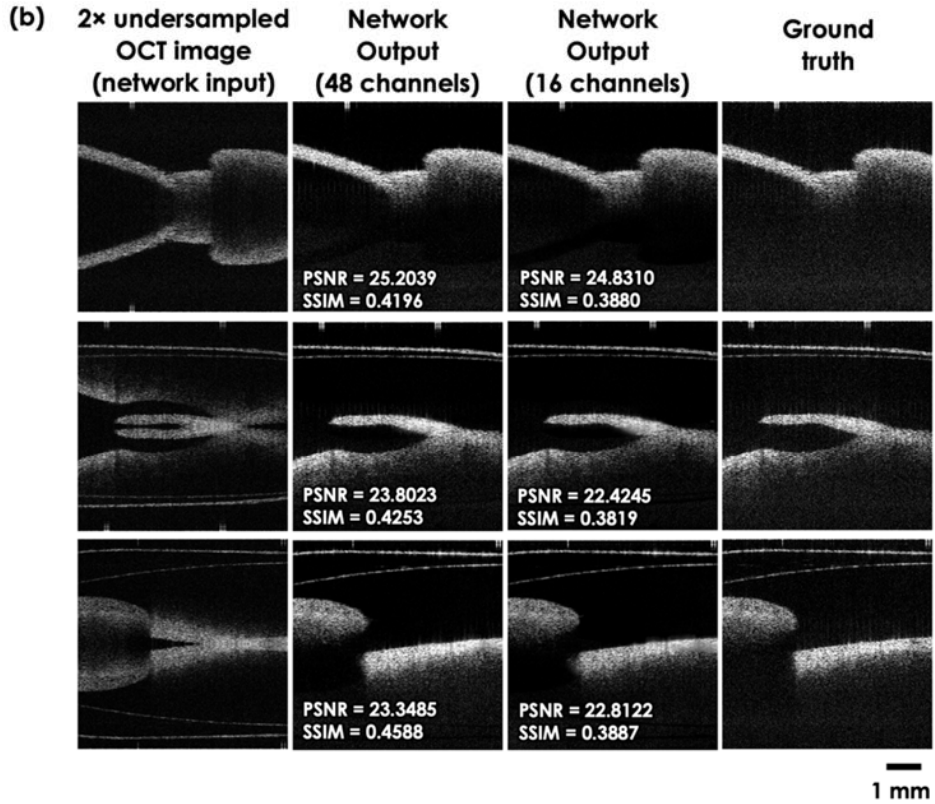

**Fig. 5 DL-OCT inference time as a function of the B-Scan batch size for blind testing.** (a) With increasing batch size, the average inference time per B-Scan (512 A-lines) rapidly decreases owing to the parallelizable nature of the neural network computation. The average inference time converged to ~6.73 ms per B-Scan for a batch size of 128. If the number of channels in the neural network's first layer is reduced from 48 down to 16, the average inference time further improved to ~1.69 ms per B-Scan. Our GPU memory size limited further reduction of the average inference time of DL-OCT. All inference times were obtained by averaging 1000 independent runs, computed on a desktop computer (see Materials and Methods). (b) Sample fields-of-view are shown for network input, network output (using 48 channels vs. 16 channels in the first layer) and ground truth images. PSNR and SSIM values are also displayed for each one of these fields-of-view.

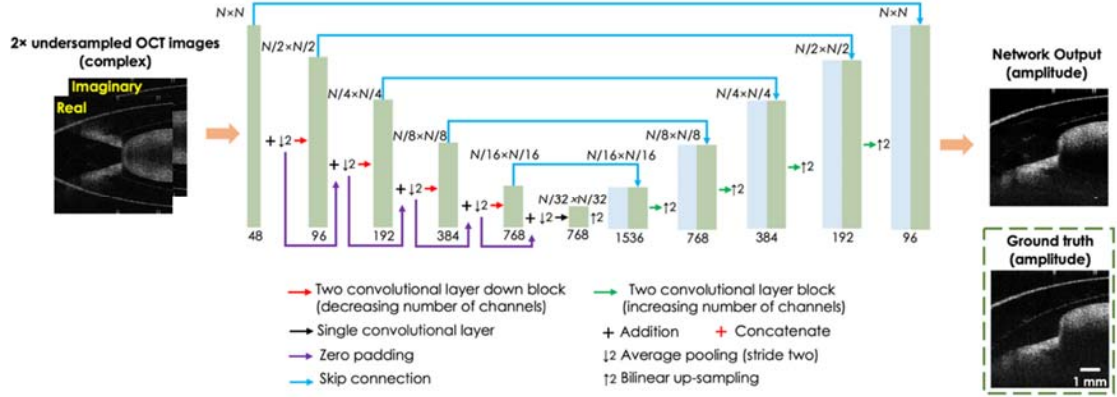

**Fig. 6 Network architecture of the encoder-decoder used in DL-OCT framework**.

Finally, we explored to see if DL-OCT image reconstruction framework can be extended to 3× undersampled spectral data per A-line. For this, we used the same neural network architecture as before, which was this time trained with input SS-OCT images that exhibited even more extensive spatial aliasing since for every spectral measurement data point that is kept, 2 neighboring wavelengths were dropped out, resulting in 427 spectral data points contributing to an A-line, whereas the ground truth images of the same samples had 1280 spectral measurements per A-line. The blind inference results of this DL-OCT network for 3× undersampled spectral data are reported in Fig. 7, which also shows, for comparison, the output images of the former DL-OCT network that was trained using 2× undersampled spectral data. This comparison in Fig. 7 reveals that, while DL-OCT can successfully process 3× undersampled spectral data with decent image reconstructions at its output, it also starts to exhibit some spatial artifacts in its inference when compared with the ground truth images of the same samples (see e.g., the red marks in Fig. 7). Furthermore, its reconstruction image quality is relatively reduced in terms of PSNR and SSIM metrics (see Fig. 7) compared with the images reconstructed by DL-OCT network that was trained using 2× undersampled spectral data.

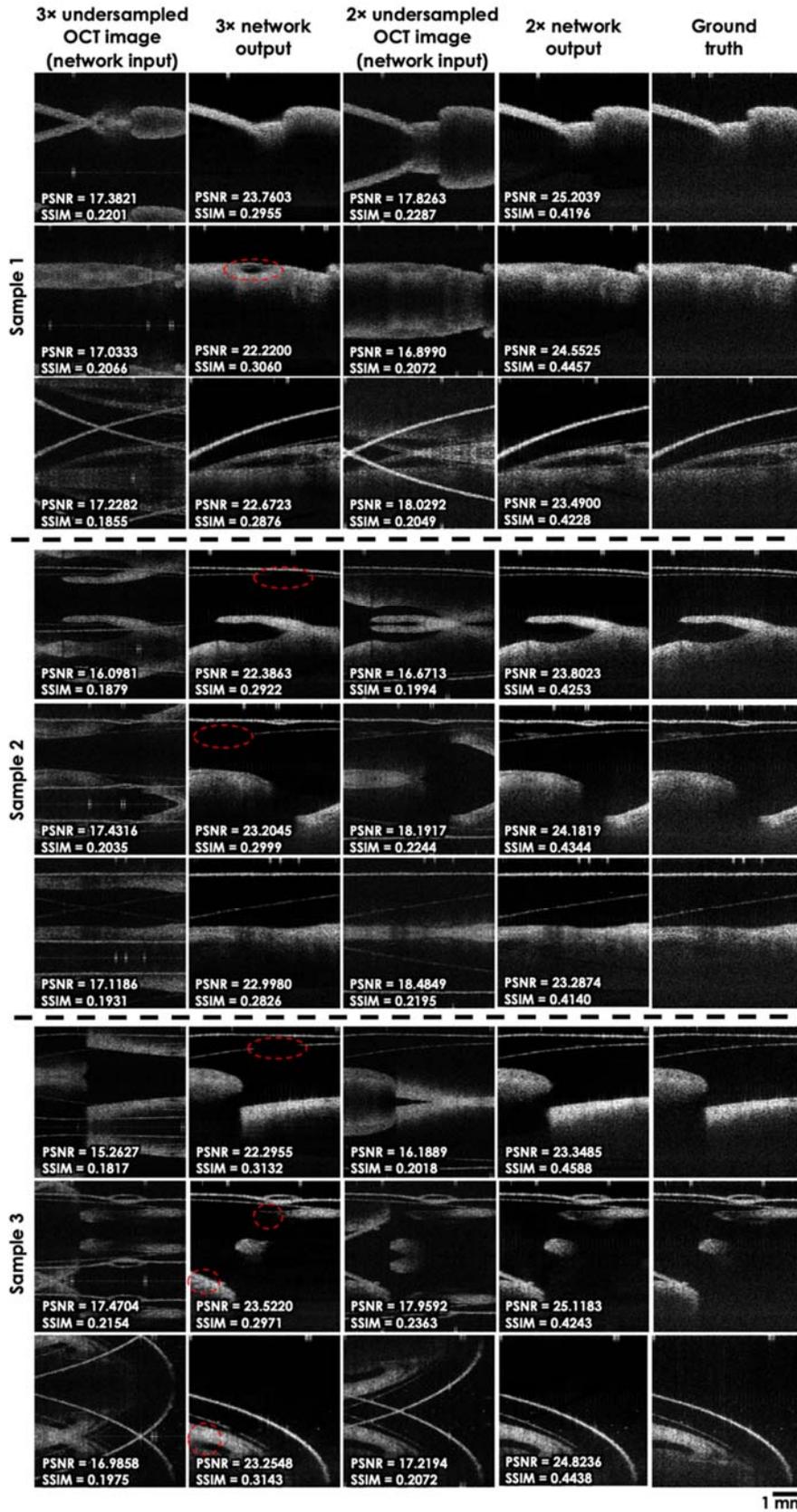

**Fig. 7 Comparison of DL-OCT blind testing results using 3× undersampled and 2× undersampled input spectral data.** Three mouse embryo samples (never seen by either of these DL-OCT networks) are imaged for

blind testing. PSNR and SSIM values are also displayed for each one of these fields-of-view. Ground truth images used 1280 spectral data points per A-line, whereas 2× and 3× DL-OCT networks used 640 and 427 spectral data points per A-line, respectively.

**Conclusion**

We demonstrated the ability to rapidly reconstruct SS-OCT images using a deep neural network that is fed with undersampled spectral data. This DL-OCT framework, with its rapid and parallelizable inference capability, has the potential to speed up the image acquisition process for various SS-OCT systems without the need for any hardware modifications to the optical set-up. Although the efficacy of this presented framework was demonstrated using an SS-OCT system, DL-OCT can also be used in various spectral domain OCT systems that acquire spectral interferometry data for 3D imaging of samples.

**Materials and Methods**

*Data acquisition*

All the animal handling and related procedures were approved by the Baylor College of Medicine Institutional Animal Care and Use Committee and adhered to its animal manipulation policies. Timed matings of CD-1 mice were setup overnight. The presence of a vaginal plug was considered 0.5 days post coitum (DPC). At 13.5 DPC, embryos (N=8) were dissected out of the mother and immediately prepared for OCT imaging. Special care was taken to ensure that the yolk sac was not damaged during dissection. The embryos were immersed in Dulbecco's Modified Eagle Media (DMEM) in a standard culture dish and imaged with the SS-OCT system (OCS1310V2, Thorlabs Inc., NJ, USA). The OCT system had a central wavelength of ~1300 nm, a sweep range of ~100 nm, and an incident power of ~12 mW. The axial and transverse resolutions of the system have been characterized as ~12 μm and ~10 μm, respectively, in air. More details on the performance of the OCT system can be found in previous work[45]. In this work, a sample area of 12 mm × 12 mm × 6 mm (X, Y, Z) was imaged. Each raw A-scan consisted of 1280 spectral data points that were sampled linearly in the wavenumber domain by a k-clock on the OCT system. 3D imaging was performed by raster scanning the OCT beam across the sample with a pair of galvanometer-mounted mirrors. Each B-scan consisted of 5000 A-scans, and each sample volume consisted of 1000 B-scans.

*Image processing*

After the data acquisition, the raw OCT fringes were processed using 2× down-sampling (by eliminating every other spectral data point), followed with zero interpolation to generate the 2× spectrally undersampled SS-OCT reconstruction (which is used as the network input). Reconstruction of the target SS-OCT image (ground truth) from the raw spectral data was performed using multiple steps. First, to decrease effect of sharp transitions and spectral leakage, each raw A-scan was windowed with a Hanning window. Next, the filtered fringes were processed by a Fast Fourier Transform (FFT) to get complex OCT data. Then, the norm of the complex vector was converted to dB scale, and the complex conjugate was discarded. A background subtraction step was performed by subtracting the mean of all the A-scans in each OCT volume from each A-scan. The resulting B-scans (after the background subtraction and windowing) were utilized as the network training targets (ground truth).

For 2× down-sampling of the measured spectral data points, the even elements of the acquired spectrum for each A-line were removed. For 3× down-sampling results reported in Fig. 7, two successive spectral measurements were eliminated, in a repeating manner, for each spectral data point that was kept. Next, zeros were interpolated in the exact same positions, where the spectral data points were removed. Then, mean of the zero interpolated spectral data was subtracted out before applying the FFT function. Both the real and imaginary parts of the down-sampled OCT complex data, resulting from the FFT, were kept as input data for the network. Each pair of input and ground truth images were normalized such that they have zero mean and unit variance before they were fed into the DL-OCT network.

*DL-OCT network architecture, training, and validation*
For DL-OCT, we used a modified U-net architecture[46] as shown in Fig. 6. Following the processing of the down-sampled OCT reconstructions and regular OCT images (ground truth images, using all the spectral data points), the resulting volumetric images were partitioned into patches of 640×640 pixels, forming training image pairs (B-scans); all blank image pairs (without sample features) were removed from training. The training loss function was defined as:

$$l = L_1 \left\{ z_{label}, G(x_{input}) \right\} \quad (1)$$

where G(·) refers to the output of the neural network, $z_{label}$ denotes the ground truth SS-OCT image without undersampling, and $x_{input}$ represents the network input. The mean absolute error, $L_1$ norm, was used to regularize the output of the network and ensure its accuracy.

The modified version of the U-net architecture is shown in Fig. 6, which has five down-blocks followed by five up-blocks. Each one of the down-blocks consists of two convolution layers and their activation functions, which together double the number of channels. A max pooling layer with a stride and kernel size of two is added after the two convolution layers to down sample the features. The up-blocks first upscale the output of the center layer using bilinear interpolation by a factor of two. And then two convolution layers and their activation functions, which decrease the number of channels by a factor of two, are added after the upscaling. Between each one of the up- and down-sampling blocks of the same level, a skip connection concatenates the output of the down-blocks with the up-sampled images, enabling the features to be directly passed at each level. After these down- and up-blocks, a convolution layer is used to reduce the number of channels to one, which corresponds to the reconstructed output image, approximating the ground truth OCT image.

Throughout the U-net structure, the convolution filter size is set to be 3×3; the output of these filters is followed by a Leaky ReLU (Rectified Linear Unit) activation function, defined as:

$$\text{Leaky ReLU}(x) = \begin{cases} x & \text{for } x > 0 \\ 0.1x & \text{otherwise} \end{cases} \quad (2)$$

The learnable variables were updated using the adaptive moment estimation (Adam[47]) optimizer with a learning rate of $10^{-4}$. The batch size for the training was set to be 3.

*Quantitative metrics*
PSNR is defined as:

$$\text{PSNR} = 10 \times \log_{10} \left( \frac{\text{MAX}_\mathbf{I}^2}{\text{MSE}} \right) \quad (3)$$

where $\text{MAX}_\mathbf{I}$ is the maximum possible pixel value of the ground truth image. MSE is the mean squared error between the two images being compared, which is defined as:

$$\text{MSE} = \frac{1}{n^2} \sum_{i=0}^{n-1} \sum_{j=0}^{n-1} \left[ \mathbf{I}(i,j) - \mathbf{K}(i,j) \right]^2 \quad (4)$$

where **I** is the target image, and **K** is the image that is compared with the target.

SSIM is defined as:

$$\text{SSIM}(a,b) = \frac{(2\mu_a\mu_b + C_1)(2\sigma_{a,b} + C_2)}{(\mu_a^2 + \mu_b^2 + C_1)(\sigma_a^2 + \sigma_b^2 + C_2)} \tag{5}$$

where $\mu_a$ and $\mu_b$ are the mean values of $a$ and $b$, which represent the two images being compared, $\sigma_a$ and $\sigma_b$ are the standard deviations of $a$ and $b$, $\sigma_{a,b}$ is the cross-covariance of $a$ and $b$, respectively, and $C_1$ and $C_2$ are constants that are used to avoid division by zero. Note that both PSNR and SSIM metrics can be affected by background noise in an OCT image. Therefore, to compute these two metrics we used the network output and target (ground truth) images that are over the noise level (70 dB in our SS-OCT system) and then converted them into gray scale with a range from 0 to 1, using double precision.

*Implementation details*

The network was implemented using Python version 3.6.0, with TensorFlow framework version 1.11.0. Network training was performed using a single NVIDIA GeForce RTX 2080Ti GPU (Nvidia Corp., Santa Clara, CA, USA) and testing was performed using a desktop computer with 4 GPUs (NVIDIA GeForce RTX 2080Ti). The training process using ~20K image pairs (640 A-lines in each image) took about 18 hours. DL-OCT inference times as a function of the batch size are reported in Fig. 5.